# Software Defect Prediction using Adaptive Differential Evolution-based Quantum Variational Autoencoder-Transformer (ADE-QVAET) Model

**Seshu Babu Barma**
**Apple**
barma_sb@apple.com

**Mohanakrishnan Hariharan**
**Apple**
m_hariharan@apple.com

**Satish Arvapalli**
**Apple**
sarvapalli@apple.com

*Abstract*- **An AI-ML-powered quality engineering approach uses AI-ML to enhance software quality assessments by predicting defects. Existing ML models struggle with noisy data types, imbalances, pattern recognition, feature extraction, and generalization. To address these challenges, we develop a new model, Adaptive Differential Evolution (ADE) based Quantum Variational Autoencoder-Transformer (QVAET) Model (ADE-QVAET). ADE combines with QVAET to obtain high-dimensional latent features and maintain sequential dependencies, resulting in enhanced defect prediction accuracy. ADE optimization enhances model convergence and predictive performance. ADE-QVAET integrates AI-ML techniques such as tuning hyperparameters for scalable and accurate software defect prediction, representing an AI-ML-driven technology for quality engineering. During training with a 90% training percentage, ADE-QVAET achieves high accuracy, precision, recall, and F1-score of 98.08%, 92.45%, 94.67%, and 98.12%, respectively, when compared to the Differential Evolution (DE) ML model.**

## I. INTRODUCTION

Software testing in enterprise environments faces challenges due to complex data and business requirements. Quality Engineers (QEs) spend 30-40% of their time performing manual test execution. Test Execution involves a manual approach of identifying and logging defects. This manual approach causes significant bottlenecks in software development lifecycles, particularly in complex enterprise systems such as cloud, microservices, IoT, and AI implementations, where intricate business logic and technical dependencies create exponential complexity.

Large Language Models (LLMs) and Generative Artificial Intelligence have opened new possibilities for automating quality engineering processes. However, traditional approaches face limitations like hallucination, context-poor generation, and loss of critical business relationships during retrieval.

These limitations become particularly pronounced in enterprise software testing, where maintaining traceability between requirements, defects, and business logic is paramount for regulatory compliance and quality assurance.

Unlike prior approaches that treat Differential Evolution (DE), Variational Autoencoders (VAE), or Transformers in isolation, our proposed ADE-QVAET framework integrates these into a unified model. ADE provides dynamic hyperparameter tuning beyond traditional DE, QVAE extracts richer high-dimensional latent features beyond classical VAEs, and the Transformer layer preserves sequential dependencies better than standard encoders. Together, these innovations enhance quality in defect prediction accuracy and scalability.

### A. Problem Statement

Current software testing methodologies face several critical challenges:

- **Software Testing Cost**: Routine testing practices, such as manual functional testing and other methods, need significant human involvement, which drives up costs, slows testing duration, and creates a higher possibility of human errors

- **Defect Prevention**: Testing methods depend on finding errors once development ends, and ignore potential failures that could be prevented earlier

- **Model accuracy**: Software test quality and security get better through AIML technology to predict defects and build tests; however, the accuracy of existing models is less than 80%

- **Defect Identification**: Historical testing knowledge remains trapped in individual expertise to identify defects, rather than organizational assets

### B. Research Contributions

This research introduces an AI-powered quality engineering approach that enhances software defect prediction using the ADE-QVAET model.

The Adaptive Noise Reduction and Augmentation (ANRA) framework improves results by reducing noise and balancing defect instances. The QVAET model detects accurate defects by extracting high-dimensional latent features while preserving sequential dependency. The ADE algorithm optimizes hyperparameters dynamically and adjusts the scaling factor and crossover rate based on the evolving performance of candidate solutions.

The ADE-QVAET model unifies ADE optimization with the QVAET model structure. ADE-QVAET provides a balance between exploration and exploitation to deliver defect analysis, minimize inaccurate results, and optimize software quality monitoring.

### C. Methodology Overview

Our approach represents a systematic evolution from ADE and QVAET into ADE-QVAET. This progression demonstrates measurable accuracy improvements by high-dimensional feature extraction and transformer-based architecture to detect dependencies.

The core innovation lies in improving software defect prediction models that encounter obstacles from unbalanced/ noisy data, inadequate feature extraction, and weak hyperparameter adjustment, resulting in poor prediction accuracy.

### D. Paper Organization

The structure of the research paper is as follows: Section II describes earlier studies on software defect prediction. Section III explains the proposed methodology for the software defect prediction strategy. Section IV reviews the mathematical modeling of the ADE-QVAET model. Section V offers a discussion of performance evaluation and implementation. Section VI concludes the analysis and provides some potential directions for further investigation.

## II. RELATED WORK

### A. Literature Review

Khalid et al. [1] combined machine learning with feature selection and K-means clustering to improve defect prediction accuracy, but at the cost of significant computational overhead. Tang et al. [2] proposed the Adaptive Variable Sparrow Search Algorithm (AVSSA), which enhanced the global search capability and handled imbalanced data effectively; however, its ensemble complexity increased the runtime. Khleel et al. [3] integrated CNN and GRU with SMOTE-Tomek sampling to address class imbalance and enhance prediction accuracy, although the approach required substantial computation and synthetic data generation. Mehmood et al. [4] advanced a Feature Selection-Based ML method across NASA datasets, improving classification precision but risking information loss when features were inappropriately discarded.

While prior studies have demonstrated progress in software defect prediction, they are impacted by computational complexity, weak handling of noisy and imbalanced datasets, and limited generalization to evolving software projects. Our ADE-QVAET framework addresses these gaps by combining adaptive noise reduction, high-dimensional latent feature extraction, and dynamic hyperparameter tuning, delivering both scalability and accuracy. To maintain focus, we streamline references to the most recent and domain-relevant works, omitting peripheral studies (e.g., WEKA tools, logistic regression in education) that do not directly inform this research direction.

### B. Challenges

i)      The prediction model's accuracy suffers from uncleaned data that also contains negligible amounts of redundant information in datasets and requires effective cleaning and balancing to prevent performance degradation [1].

ii)      Deep learning models tend to develop overfitting behaviors when they receive restricted or noisy data during their training operations. Training models that absorb irrelevant data details during learning processes causes them to perform inadequately when presented with unknown data. The accuracy of defect predictions is reduced when the model encounters new or changing software projects [2].

iii)      Deep learning models represent black-box systems because experts cannot easily understand the mechanisms that lead to specific predictions from the models. Additional information from the training data is ignored, leading quality engineers not to trust their use within the software development procedure [4].

iv)      Training together with inference processes of AI models, specifically deep learning architectures, requires major computing hardware resources. The long training durations of these models create bottlenecks for quick implementation when new software needs detection or during codebase updates [5].

v)      Current systems for defect prediction fail to integrate with widely used software development platforms, which include both version control systems (e.g., Git) and CI/CD pipelines. The lack of a proper connection between these tools causes delayed predictions of defects. A direct integration with popular software development tools serves as a foundation for obtaining real-time quality feedback and achieving steady enhancements of software quality [6].

## III. PROPOSED METHODOLOGY FOR SOFTWARE DEFECT PREDICTION USING ADE-QVAET MODEL

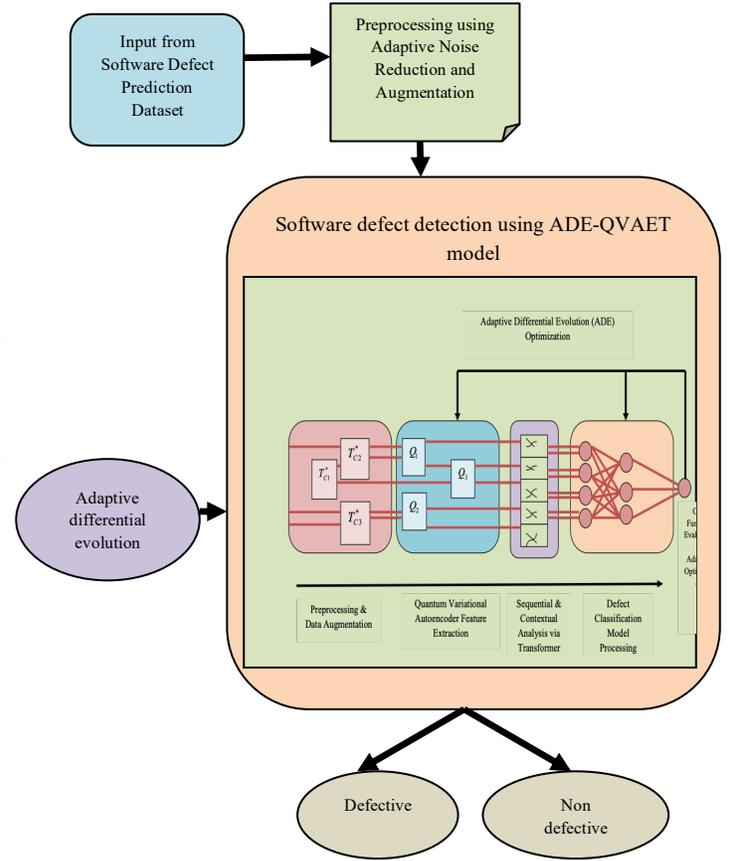

Fig. 1. Proposed methodology for software defect prediction

*A. Input collection from Software Defect Prediction Dataset:*

The primary input dataset originates from the software defect prediction dataset https://www.kaggle.com/datasets/semustafacevik/software-defect-prediction. This dataset includes static code attributes, maintainability index, cyclomatic complexity, lines of code, and code churn features. These performance indicators provide insights into defect-prone modules, enabling the model to identify patterns related to software quality levels. The dataset contains both working examples and faulty outcomes, which facilitates the model's learning process.

$$I = \sum_{b=1}^{n} T_C$$

Here, $I$ denotes the dataset, $T_C$ denotes the number of metrics present in the dataset, with values ranging from 1 to $n$.

*B. Preprocessing using Adaptive Noise Reduction and Augmentation*

Preprocessing raw input data is crucial for reliable defect prediction. The software defect prediction dataset (Kaggle) faces performance issues due to noise, inconsistent data, and unequal distribution of defect and non-defect instances. The ANRA framework addresses these issues by filtering unnecessary information and generating synthetic data to balance the class imbalance. This preprocessing step ensures data integrity, improving model accuracy and reliability for defect prediction.

$$I^{'} = \sum_{b=1}^{n} T_C^{*}$$

The preprocessed dataset following the ANRA Framework noise reduction and augmentation process is denoted a $I^{*}$ Field. The software metrics $T_C^{*}$ result from data cleaning, reduced redundancy, and implemented synthetic data generation methods, while $n$ representing the original metric count.

*C. Quantum Variational Autoencoder-Transformer model for software defect prediction:*

Data preprocessing provides refined software metrics, which serve as model input to QVAET. Through its quantum QVAE operation, the component identifies sophisticated multi-dimensional latent patterns embedded in the input metrics to help the model understand data patterns. Through transformer-based architecture, the model examines features to enable sequential dependencies with contextual relationships between software metrics to enhance its capability for predicting defect-prone modules.

$$G = QVAET\left(T_C^{*}\right)$$

The quantum variational autoencoder extracts latent features from the input $T_C^{*}$, which serves as the input. The extracted features consist of high-dimensional structures that detect intricate data patterns.

This architecture detects sequence connections and contextual relationships between metrics in software systems. Self-attention Mechanisms determine metric relevance for other time-based or context-related metrics in software projects. By recognizing software attributes, the model understands their collective influence on defect likelihood within modules. Sequential data processing and context consideration improve defect-prone module detection by identifying hidden patterns beyond normal inspection methods.

$$Q = Transformer\left(G\right) \tag{4}$$

The transformer-based architecture produces output $Q$ by processing $G$ latent features while identifying important sequences and contextual associations between software metrics.

The processed features $Q$ generated from the Transformer advance to the prediction layer of the model. The prediction layer evaluates the learned patterns and Relationships to determine the probability that a software module contains defects. The model applies processed Features to decide whether the module contains defects or not. Here $\hat{z}$ is the final prediction (defect-prone or non-defect-prone) that reflects the inference output based on the features. Using high-level features and relationships learned during the previous steps, the model makes an informed decision, which ultimately helps in identifying high-risk software modules and aids in the defect prediction process, leading to enhanced software quality assurance.

$$\hat{z} = prediction\left(Q\right) \tag{5}$$

## IV. PROPOSED ADAPTIVE DIFFERENTIAL EVOLUTION ALGORITHM

The ADE algorithm, an optimization technique, adapts the hyperparameters of machine learning models during training to enhance their performance. It extends the traditional DE algorithm, which is a population-based optimization method. ADE incorporates adaptive mechanisms to adjust DE parameters, such as the scaling factor and crossover rate, based on the evolving level of the population during the optimization process.

Fig. 3. The ADE-QVAET step-by-step process

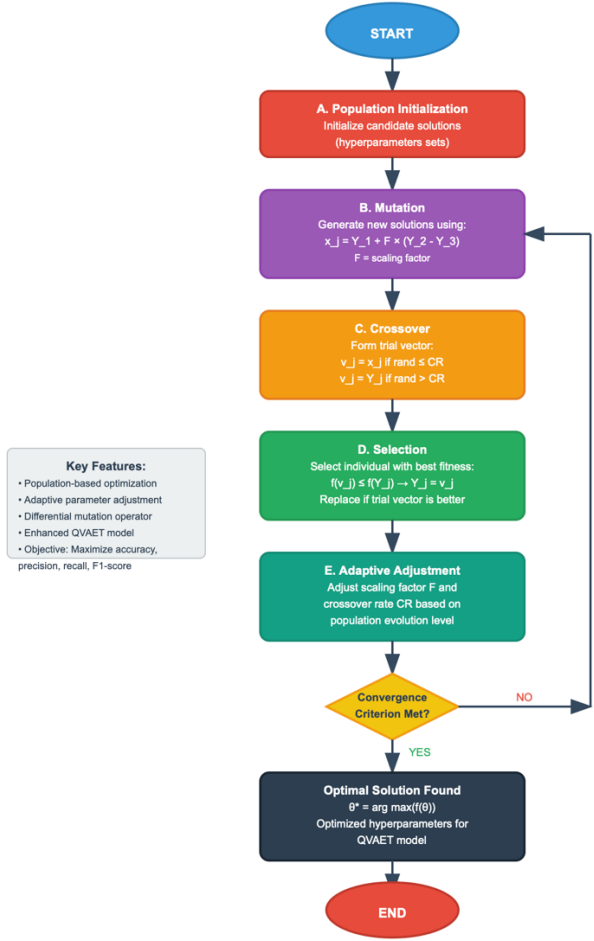

### A. Population Initialization and Mutation

ADE initializes a population of candidate solutions (sets of hyperparameters). The mutation operator generates new solutions, and individuals are selected and combined using a differential formula to increase diversity. The mutation step explores the search space by perturbing candidate solutions.

$$x_j = Y_s 1 + F \times (Y_s 2 - Y_s 3)$$

The mutated individual $x_j$ serves as the candidate solution, while the population selection generates three random individuals $Y_s 1, Y_s 2, Y_s 3$. $F$ The scaling factor that determines the step size control of the mutation.

### B. Crossover

After mutation, the trial vector is formed by the combination of the current candidate and the mutated solution.

$$v_j = \begin{cases} x_j & if\ rand \le CR \\ Y_j & if\ rand > CR \end{cases}$$

The mutation process relies on the crossover rate $CR$ & $rand$, which is a random number between 0 and 1. This random number determines whether to accept the mutated solution or retain the original one.

### C. Selection

The next step is to generate the trial vector and select the individual with the best fitness value. If the trial vector performs better than the current candidate, it replaces the existing individual in the population.

$$f(v_j) \le f(Y_j) \Rightarrow Y_j = v_j$$

### D. Adaptive Adjustment

The primary contribution of ADE lies in its ability to adaptively adjust the scaling factor and crossover rate during the optimization process. This adaptability is achieved through past candidate solutions to determine whether these parameters are more effective for exploring or converging in future acquisitions. If a solution fails to improve the population, the algorithm may reduce the scaling factor or modify the crossover rate to refine the search process.

### F. Convergence

It is an iterative process of mutation, crossover, and selection until the convergence criterion is met (for example, the desired number of generations or achieving satisfactory values of fitness).

The objective of ADE is to optimize (or minimize) the objective function $f(\theta)$ that assesses the quality of the model when an optimal set of hyperparameters $\theta^*$ is used. In this research, the model's performance metrics are accuracy, precision, recall, and F1-score, which are determined by the hyperparameters.

$$\theta^* = \arg\max_{\theta}(f(\theta))$$

To enhance the QVAET model, the ADE algorithm optimizes its most crucial parameters, such as the learning speed, regularization coefficients, and the total number of layers. This optimization process facilitates faster convergence and ensures that the model achieves better accuracy while maintaining its effectiveness across diverse software defect testing tasks. By automatically optimizing parameters, the model generates more accurate predictions, leading to more reliable results in identifying defect-prone software modules.

## V. RESULT AND DISCUSSION

This study developed the ADE-QVAET model to predict software defects. It was then compared with other top models to assess its performance and effectiveness.

### A. Dataset description:

1) Software Defect Prediction Dataset description [22]

The Kaggle software defect prediction dataset evaluates defect sensitivity based on code lines (LOC), cyclomatic complexity, depth of inheritance tree (DIT), coupling between objects (CBO), and other structural elements. It contains two outcomes: defective (1) and non-defective (0), suitable for binary classification tasks. The dataset covers multiple software versions, enabling researchers to study model generalization. However, the class imbalance problem affects prediction accuracy due to the infrequent occurrence of defective modules. This dataset is a fundamental resource for developing early defect detection systems that improve software quality and reliability.

*B. Performance Analysis based on TP:*

The ADE-QVAET model's effectiveness in software defect prediction across varying epochs (100, 200, 300, 400, and 500) while maintaining a TP of 90 is as follows.

Figure 2a shows remarkable accuracy levels: 74.01%, 92.34%, 90.34%, 90.11%, and 98.67%, all at a TP of 90.

Figure 2b shows the highest precision scores: 90.23%, 90.89%, 90.23%, 89.89%, and 98.67%, all at a TP of 90.

Figure 2c shows the highest recall rates: 90.44%, 90.76%, 89.45%, 88.33%, and 93.34%, all at a TP of 90.

Figure 2d shows peak f1-score values: 90.33%, 90.78%, 90.67%, 89.45%, and 98.56%, all at a TP of 90.

*C. Comparative methods*

To emphasize the accomplishments of the ADE-QVAET models, a comparison was done. This investigation employed several techniques, such as SVM [16], DT [17], RF [18], LR [19], QVA [20], and DE [21].

**1) Comparative analysis based on TP**

The ADE-QVAET model showed better performance than the DE model in predicting software defects at a TP of 90, reflecting an improvement of 7.73% and reaching a peak accuracy of 98.08%, as shown in Figure 4a.

In Figure 2b, the ADE-QVAET model displays improved predictive abilities in software defect forecasting compared to the DE model, achieving an 18.63% advantage and a top precision of 92.45% at a TP of 90.

Figure 2c reveals that the ADE-QVAET model outperformed the DE model by 4.34% in its software defect predictions, with a maximum recall of 94.67% at a TP of 90, thus exceeding earlier techniques.

Figure 2d demonstrates that the ADE-QVAET model surpasses the DE model in software defect prediction by recording an f1-score of 98.12% at a TP of 90, which is 15.63% higher than that of the DE model.

Fig. 2a. Accuracy at different epochs

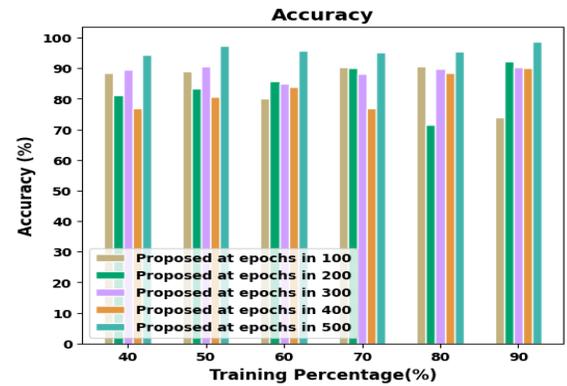

Fig. 2b. Precision at different epochs

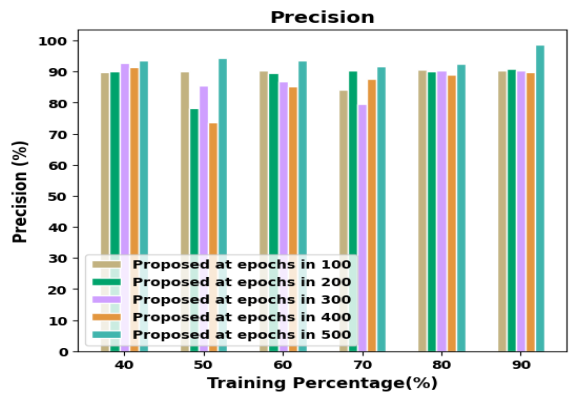

Fig. 2c. Recall at different epochs

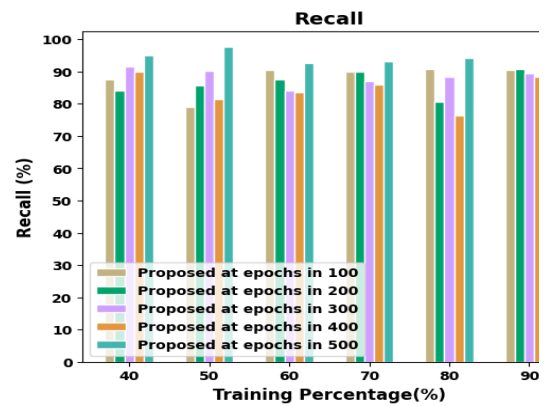

Fig. 2d. F1 score at different epochs

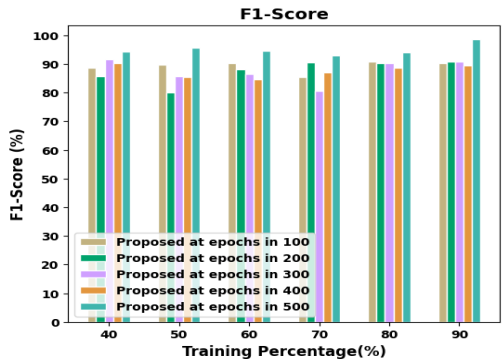

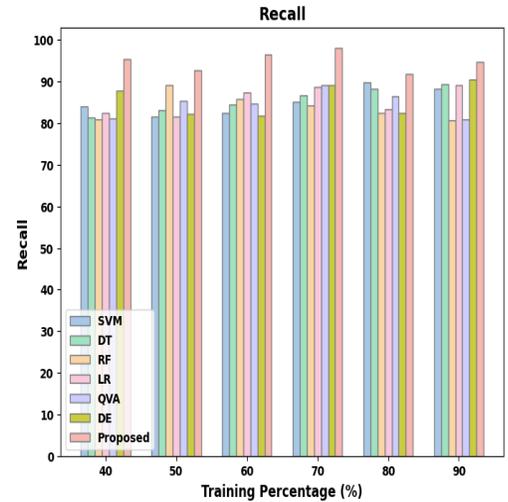

Fig 2c. Recall of ADE-QVAET against other models

Comparative Analysis based on TP 4a) Accuracy, 4b) Precision, 4c) Recall, and 4d) F1-score

Fig 2a. Accuracy of ADE-QVAET against other models

Fig 2d. F1-Score of ADE-QVAET against other models

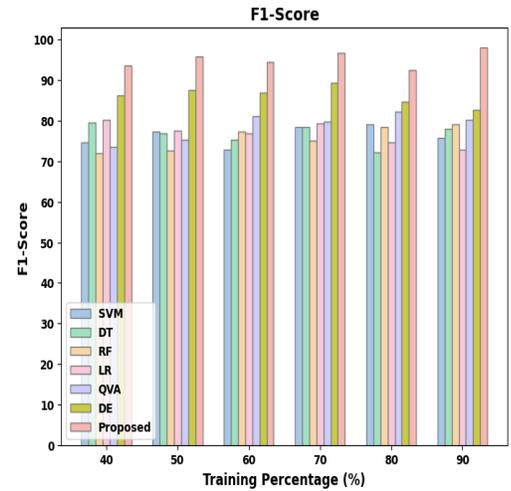

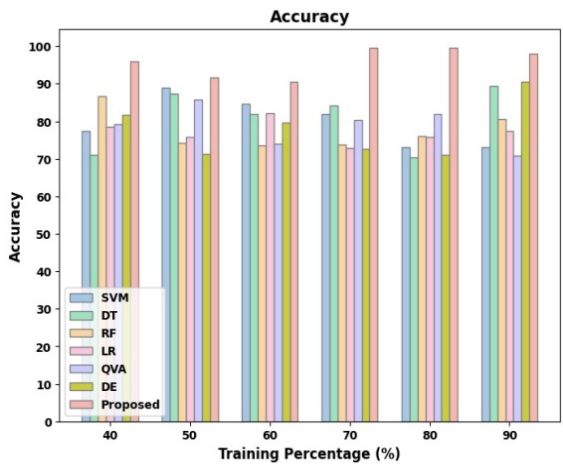

Fig 2b. Precision of ADE-QVAET against other models

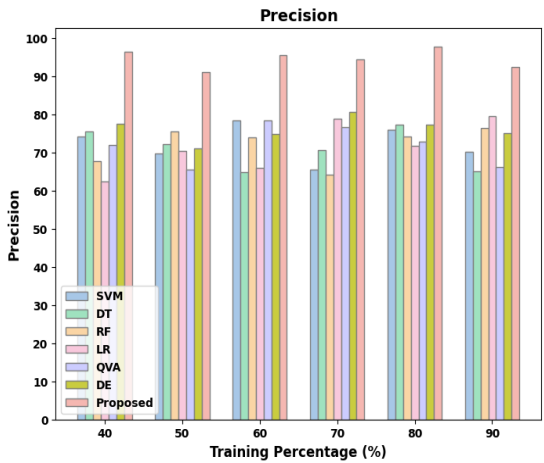

### E. Comparative discussion table

Existing models like SVM, DT, RF, and LR struggle with high-dimensional data, noise sensitivity, imbalanced datasets, and complex feature relationships, hindering software defect prediction. ADE-QVAET addresses these issues by uniting QVAE with transformer-based sequence processing for high-dimensional feature extraction. ADE hyperparameter tuning improves model convergence and accuracy, while ANRA data quality management eliminates noise and creates artificial samples for balanced class distribution. These strategies enhance performance by handling data quality issues and detecting intricate patterns for defect prediction. A comparative discussion table is presented in Table I.

| Model | TP 40/50/60/70/80/90 | | | |
|---|---|---|---|---|
| | Accuracy | Precision | Recall | F-score |
| SVM | 77.49/89.01/84.64/81.97/7 3.12/73.12 | 74.23/69.89/78.56/65.47/7 6.12/70.33 | 84.12/81.56/82.45/85.23/8 9.78/88.34 | 74.56/77.23/72.89/78.34/7 9.12/75.12 |
| DT | 71.16/87.32/82.02/84.16/7 0.41/89.4 | 75.56/72.23/64.89/70.67/7 7.34/65.23 | 81.23/83.12/84.45/86.67/8 8.23/89.45 | 79.45/76.78/75.34/78.45/7 2.12/77.89 |
| RF | 86.65/74.25/73.64/73.67/7 6.08/80.5 | 67.89/75.67/74.12/64.23/7 4.34/76.54 | 80.78/89.23/85.78/84.23/8 2.34/80.56 | 71.89/72.67/77.23/75.12/7 8.56/79.12 |
| LR | 78.64/75.82/82.24/72.79/7 5.84/77.33 | 62.45/70.45/66.12/78.99/7 1.89/79.67 | 82.45/81.56/87.34/88.67/8 3.45/89.12 | 80.23/77.56/76.89/79.34/7 4.67/72.78 |
| QVA | 79.12/85.7/73.99/80.28/81 .85/70.93 | 72.12/65.67/78.45/76.78/7 2.9/66.34 | 81.12/85.34/84.67/89.23/8 6.56/80.89 | 73.45/75.34/81.12/79.67/8 2.12/80.23 |
| DE | 81.65/71.25/79.64/72.67/7 1.08/90.5 | 77.56/71.23/74.89/80.67/7 7.34/75.23 | 87.78/82.23/81.78/89.23/8 2.34/90.56 | 86.23/87.56/86.89/89.34/8 4.67/82.78 |
| ADE-QVAET | 96.08/91.71/90.65/99.49/9 9.66/98.08 | 96.45/91.23/96.67/94.45/9 7.89/92.45 | 95.45/92.78/96.45/98.12/9 1.89/94.67 | 93.67/95.89/94.45/96.78/9 2.56/98.12 |

## VI. CONCLUSION

The proposed ADE-QVAET model introduces an AI-based quality engineering approach to predict software defects. The ANRA Framework initiates the process by enhancing data quality through noise reduction, redundant information removal, and synthetic data generation to balance defect and non-defect instances. The refined dataset is then processed by the QVAET model, which generates complex latent representations using the QVAE architecture. The Transformer-based architecture identifies sequential dependencies between software metrics. The ADE algorithm possesses a dynamic parameter adjustment capability, ensuring optimal convergence and enhanced defect prediction performance. This research addresses existing model limitations by providing precise defect monitoring and improving software quality. Future AI-driven testing tools can be enhanced using deep learning and reinforcement learning to predict and prevent software issues even before development.